\documentclass[%
 aip,
 sd,%
 amsmath,amssymb,
reprint,%
]{revtex4-1}

\usepackage{graphicx,subfigure}
\usepackage{dcolumn}
\usepackage{bm}

\begin{document}

\preprint{AIP/123-QED}

\title{Thermally Induced Nonlinear Optical Absorption in Metamaterial Perfect Absorbers}

\author{Sriram Guddala}
 
\author{Raghwendra Kumar}%

\author{S. Anantha Ramakrishna}

\affiliation{%
Department of Physics, Indian Institute of Technology Kanpur, 208016, India.}

\date{\today}

\begin{abstract}
A metamaterial perfect absorber consisting of a tri-layer $\mathrm{(Al/ZnS/Al)}$ metal-dielectric-metal system with top aluminium nano-disks is fabricated by laser-interference lithography and lift-off processing. The metamaterial absorber had peak resonant absorbance at 1090 nm and showed nonlinear absorption for 600ps laser pulses at 1064 nm wavelength. A nonlinear saturation of reflectance was measured to be dependent on the average laser power incident and not the peak laser intensity. The nonlinear behaviour is shown to arise from the heating due to the absorbed radiation and photo-thermal changes in the dielectric properties of aluminium. The metamaterial absorber is seen to be damage resistant at large laser intensities of $\mathrm{25~MW/cm^{2}}$.  
\end{abstract}

\pacs{Valid PACS appear here}

\maketitle

Metamaterials (MTMs), with sub-wavelength unit cells of metal-dielectric composite structures called ``meta-atoms", show unique electro-magnetic resonances excited by the incident radiation. The resonant interaction causes the MTM to have highly dispersive ``effective" medium properties. The dispersive material parameters and perfect impedance ($\mathrm{Z=\sqrt{{\mu(\omega)}/{\varepsilon(\omega)}}}$) matching can give rise to unusual optical phenomena such as negative refractive index, sub-wavelength imaging and perfect absorption, etc., which are not found in the natural materials.~\cite{MPA1} Metamaterial perfect optical absorbers (MPA)~\cite{MPA2} with narrow and broad band resonances over various bands in the electromagnetic spectrum have shown great potential for a wide range of applications such as sensors,~\cite{MPA3} imaging devices,~\cite{MPA4} and solar cells,~\cite{MPA5} etc.
 
Though the linear optical properties of MTMs have been widely studied for applications, the nonlinear optical properties have attracted attention only more recently.~\cite{NLO-MM-Review} The large local field enhancements within the MTM structures due to resonant interactions can give rise to enhanced nonlinear response. Especially, there has been great interest in parametric amplification,~\cite{Kozyrev-PA} harmonic generation,~\cite{SHG,Kim-PRB} and optical switching~\cite{Switching-OE} in split ring resonators, fishnet structures and more complex chiral MTMs.~\cite{NLO-MM-Review} Incorporating nonlinear Kerr dielectric media for switchable MTMs have also been investigated.~\cite{SAR-PRB,Kerr-OE} The intense absorption of radiation in MPAs mediated by electromagnetic or plasmonic resonances can give rise to strong photo-thermal effects. So, these effects need to be well understood to optimize the devices based on MTM or plasmonic structures. 

\begin{figure}[httb]
\includegraphics[scale=0.16]{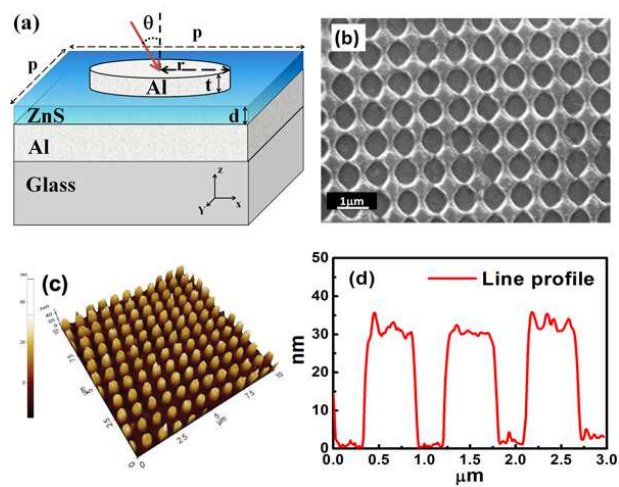}
\caption{(a) Schematic unit cell of the tri-layer (Al/ZnS/Al) MPA with structural parameters. (b) SEM top view of hole pattern photoresist on ZnS/Al films with period p= 880$\pm5$ nm and hole diameter of 560$\pm5$ nm. (c) AFM image of the top pattern Al disks after lift-off; (d) AFM image line profile shows the Al disk diameter 560$\pm5$ nm and thickness 30$\pm3$nm.}
\end{figure}
\begin{figure*}[httbt]
{\includegraphics[scale=0.26]{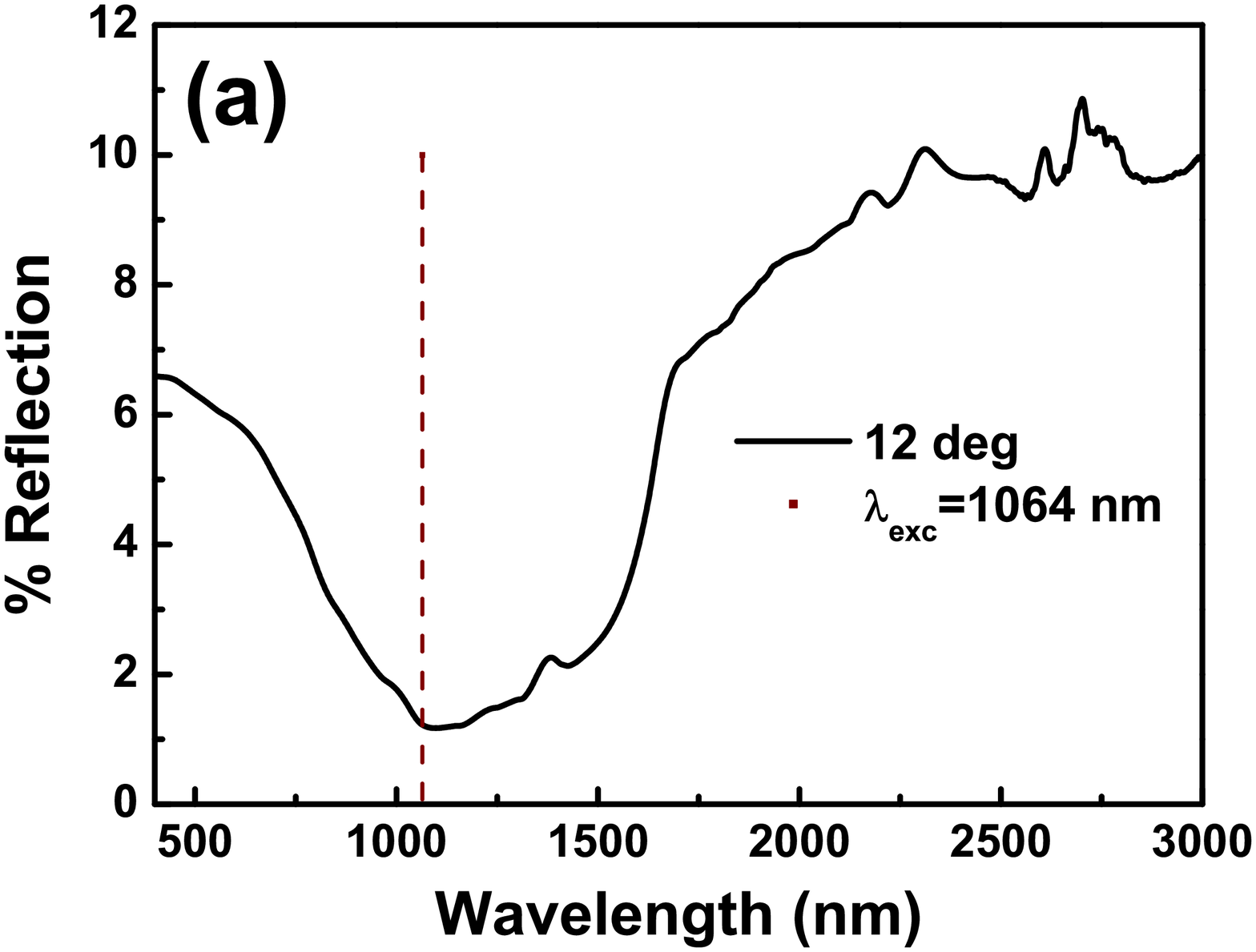}}
{\includegraphics[scale=0.25]{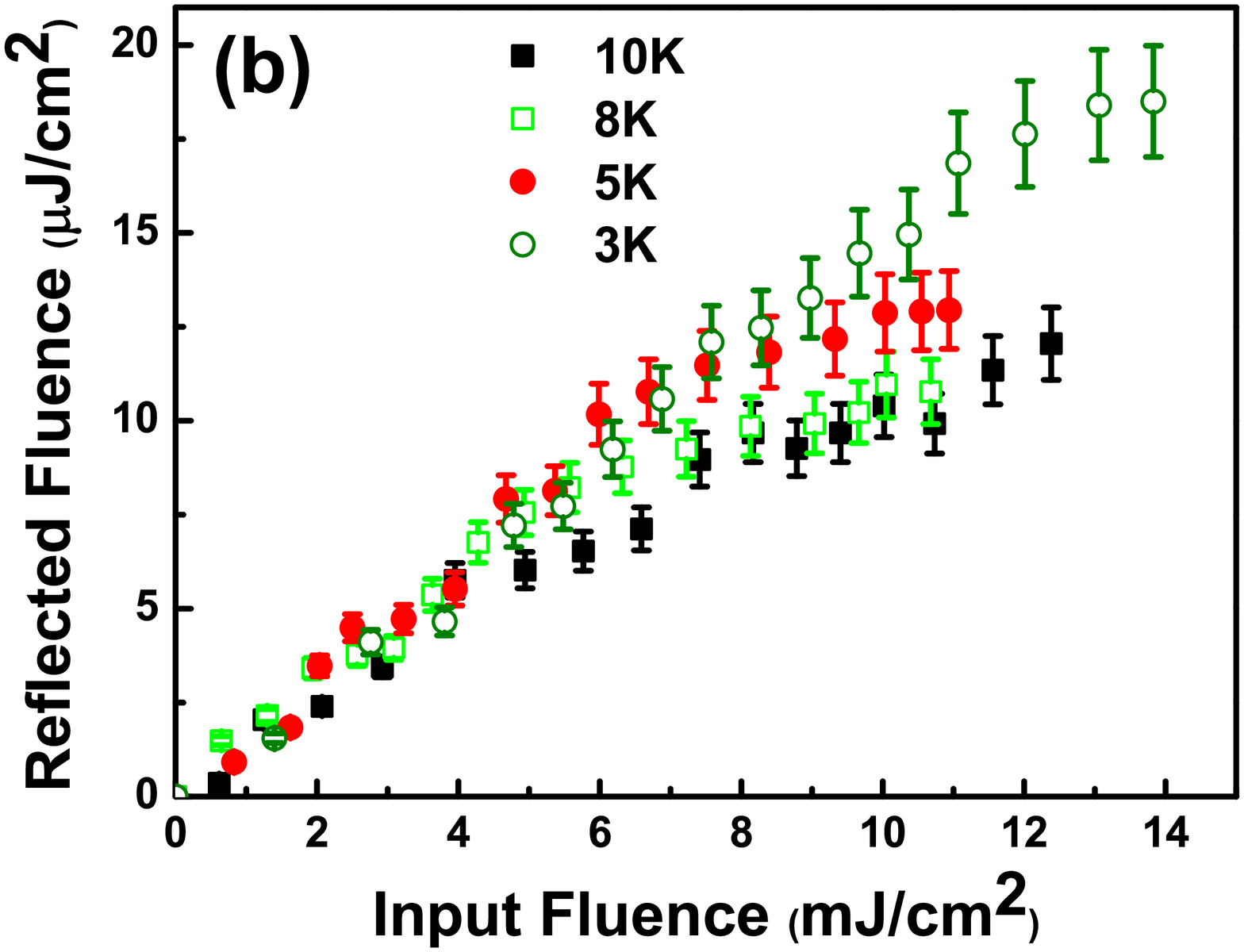}}
{\includegraphics[scale=0.26]{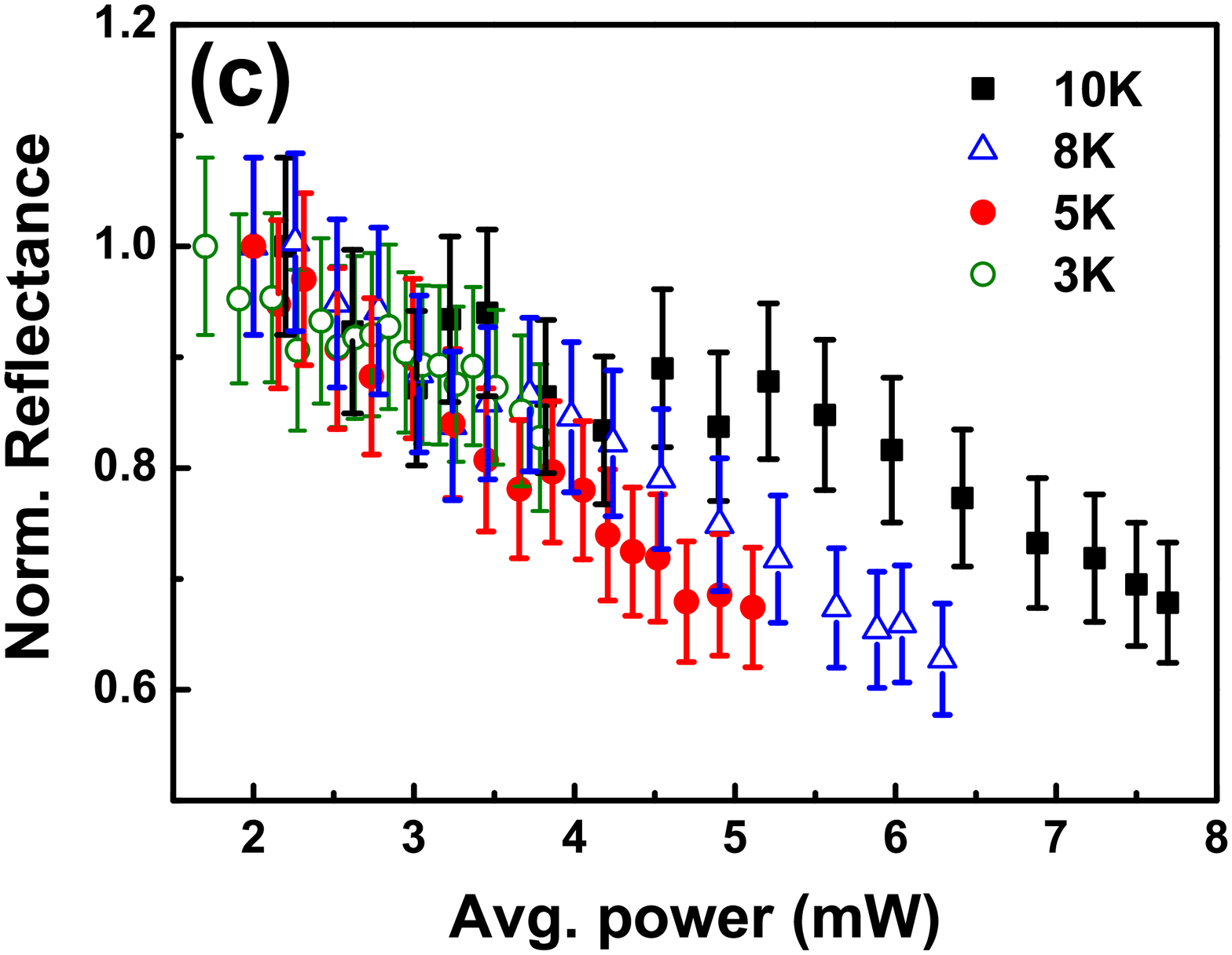}}
\caption{\label{fig:wide}(a) White light reflection spectrum of the MPA measured at $\mathrm{12^{o}}$ angle of incidence. (b) Reflected fluence verses the incident pulse fluence for each PRR (600ps, 1064 pulses). (c) Normalized reflectance plotted with respect to incident average laser power.}
\end{figure*}
In this letter, we report on thermally driven nonlinear absorption of MPAs, where the resistive heating in the structure modifies the spectral response of the MPA. We present results on nonlinear reflectance/absorbance of 600ps pulses at 1064nm from a resonant MPA consisting of a tri-layer structure of aluminium (Al) disks on ZnS/Al thin films. The constituent materials (Al, in our case) of the composite MTM change their material parameters with increase in temperature, which leads to a change in the MTM properties. We find evidence of increasing absorbance with increasing average laser power incident on the structure, indicating a purely thermal origin for the nonlinearities. These studies can have significant importance in the application of highly absorbing sensitizing layers for integrated photonic devices, thermal sensing, photo-detecting and optical imaging. We further show that MPAs can be built robustly to withstand large laser pulse interactions of several tens of $\mathrm{MW/cm^{2}}$ without any sign of damage. 

The design of the MPAs is based on resonant structures that can be simultaneously driven by both the electric and magnetic fields of radiation. This can result in resonant absorption of radiation at perfect or optimized impedance matching of the effective medium with free space. One simple design of a MPA~\cite{GD-OE,GD-JOP}(Fig. 1(a)) consists of a conducting ground plane separated by a dielectric spacer layer from the top structured metallic motif. Proper choice of the motifs and spacer layer thickness can result in simultaneous resonances for the electric and the magnetic excitations at a common frequency. Various fabrication techniques, especially, expensive and time consuming e-beam lithography~\cite{Hao-APL,PM-JAP} and focused ion beam milling techniques~\cite{He-FIB,JD-FIB} have been in wide use to fabricate motifs of MPAs with sub-micron structural features operate at NIR and visible frequencies. We have used laser interference lithography followed by lift-off processing to fabricate metallic structures with sub-micron dimensions ($<300$nm) over few $\mathrm{mm^{2}}$ areas.

The fabrication process of MPA involves three principal steps: first, an Al film (100 nm) and a thin ZnS film (65 nm) are sequentially deposited by thermal vapour deposition on a clean glass substrate. Second, a thin positive photoresist (PR) (ma-P 1205, Micro-resist Technology) film is spin coated at 3000 rpm. The PR layer was subsequently exposed to a two-beam laser interference pattern~\cite{PM-OL} followed by $\mathrm{90^{o}}$ rotation of the sample about its surface normal to obtain a 2D square lattice of holes after exposure and development as shown in Fig.1(b). Finally, a negative replica of the PR pattern is obtained by physical vapour deposition of thin Al film followed by PR lift-off by immersing in acetone for few minutes. An atomic force microscope (AFM, Park XE 70) image of the MTM in Fig. 1(c) shows the top Al disks (30 nm thickness) with a period of 880 $\pm5$ nm. Uniformity of the disk diameters (560$\pm5$ nm) and heights (30$\pm3$nm) can be noted from the AFM image line profile in Fig. 1(d). It indicates that our fabrication technique has good potential for fabricating MPAs over large areas with uniformity.

\begin{figure*}[httbt]
{\includegraphics[scale=0.27]{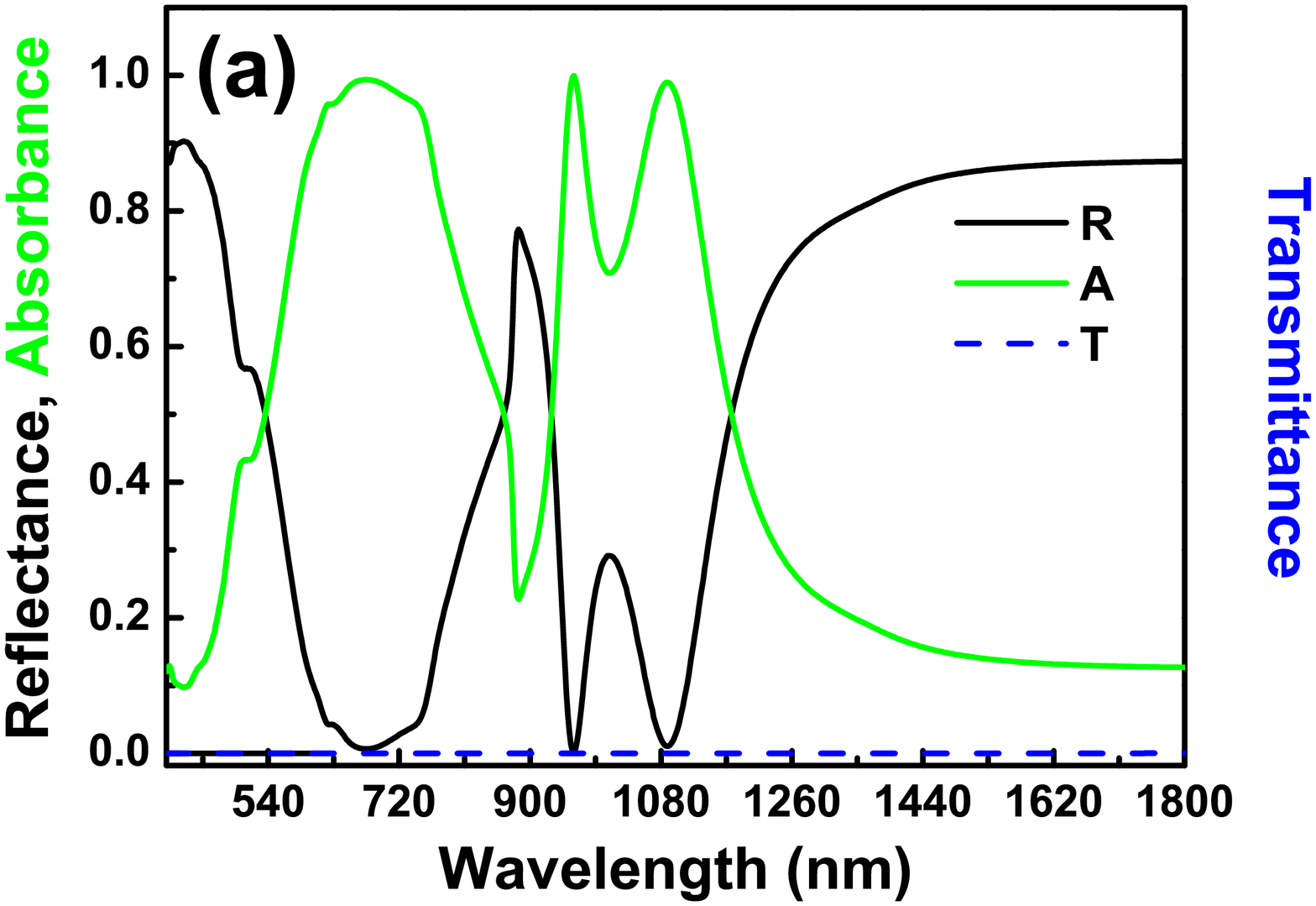}}
{\includegraphics[scale=0.24]{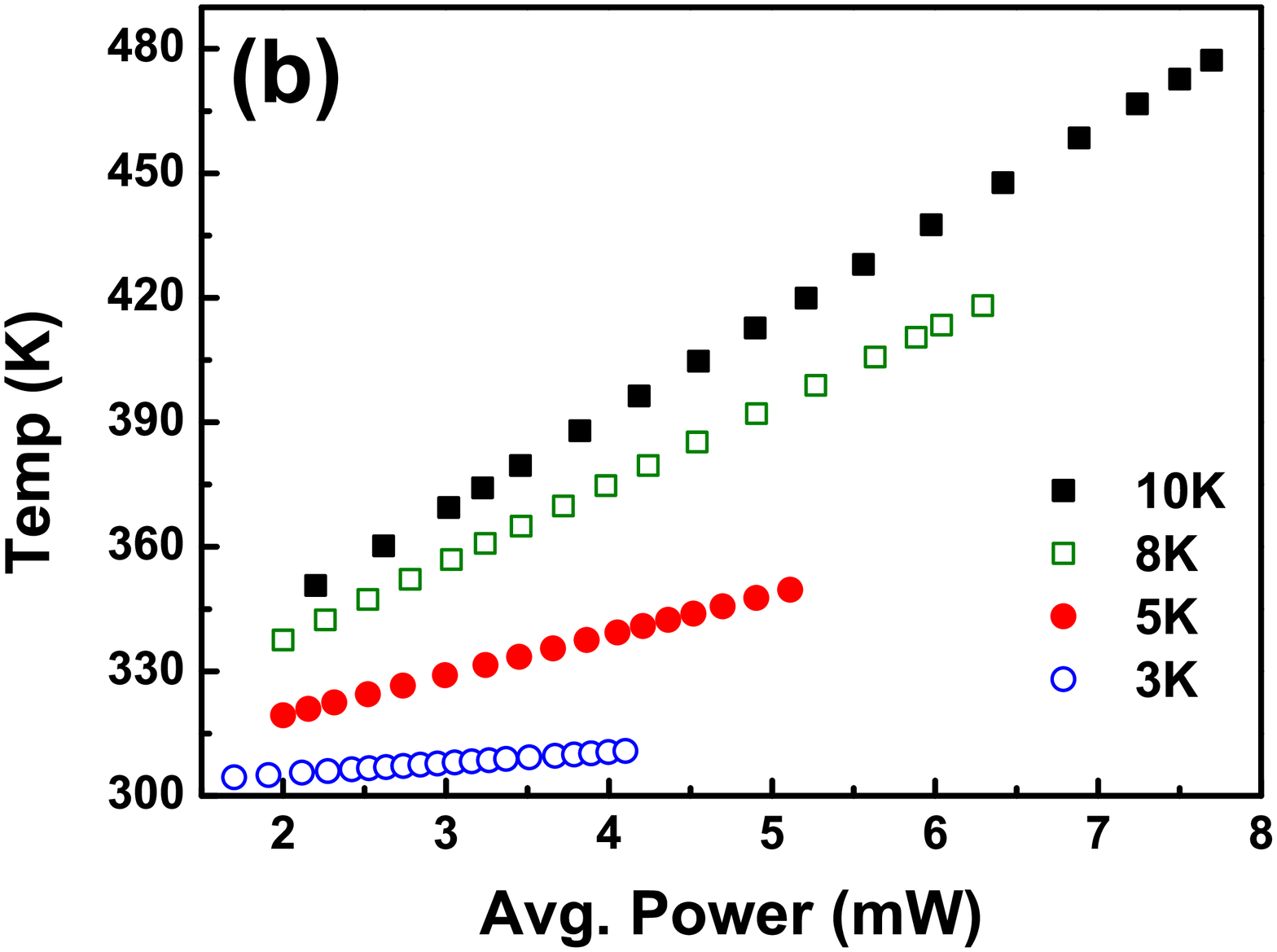}}
{\includegraphics[scale=0.24]{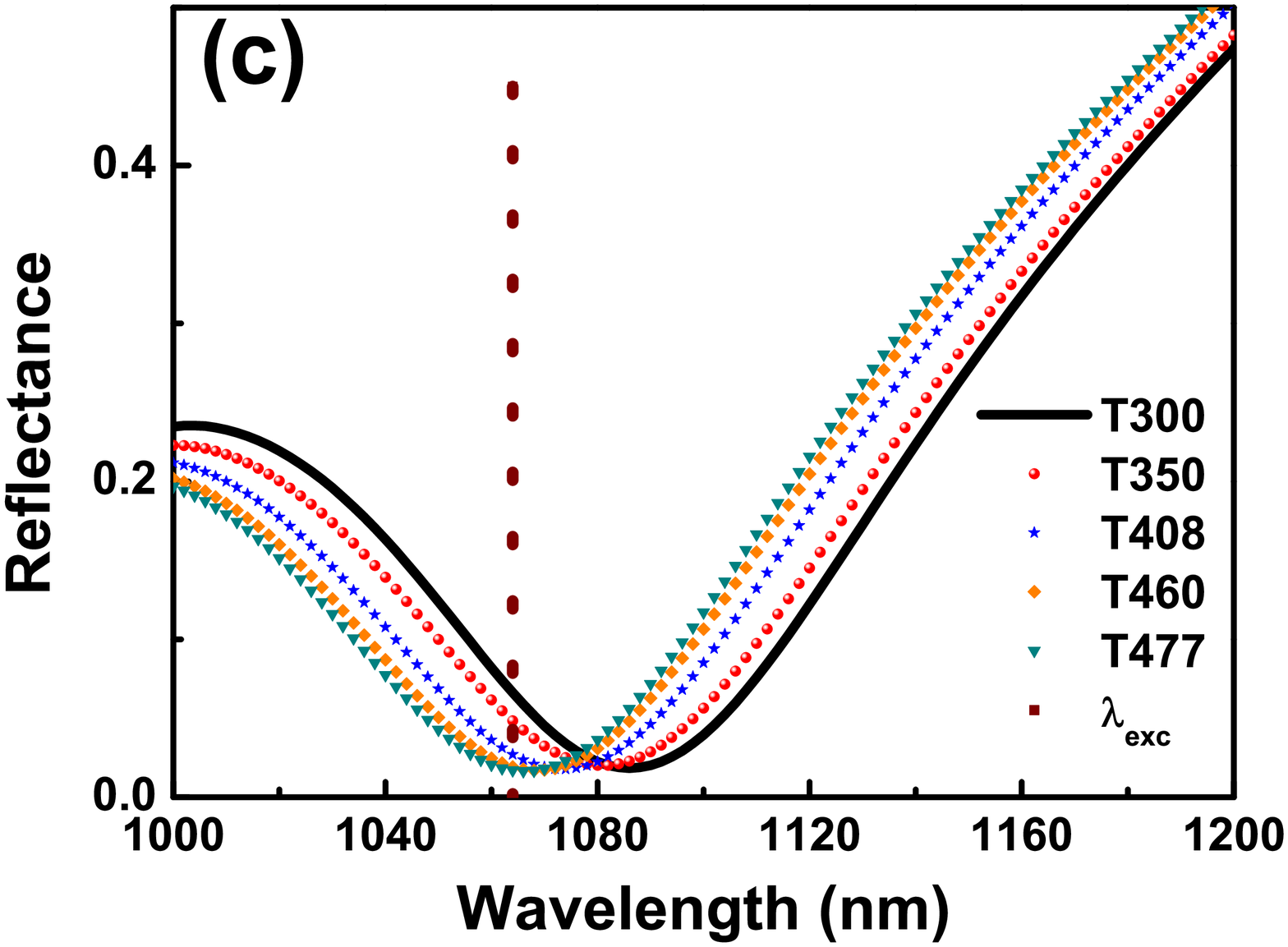}}
\caption{\label{fig:wide}{(a)Simulated spectral response of the MPA. (b)Calculated temperatures from Eq.1 for the same incident average powers at different PRRs. c) Simulated spectral response of MPA at the elevated temperatures ($\mathrm{T_{low}}$).}}
\end{figure*}

The reflection spectrum of the MPA, normalized to the reflection $(\approx100\%)$ from a thick and smooth Al film, was measured with collimated white light beam $(\mathrm{1mm\times3mm})$ at an angle of incidence of $\mathrm{12^{o}}$ with respect to the surface normal. The bottom 100 nm thick Al film, which is much more than the skin depth ($\thicksim5\mathrm{nm}$ at 1090 nm), literally makes the transmission zero. The optimized impedance  matching of the tri-layer system shows a large drop in the reflected intensity at 1090 nm within a broadband presumably due to inhomogeneous broadening of the resonances. The measured reflection spectrum [$\mathrm{A(\omega)= 1-R(\omega)-T(\omega)}$] of the MPA is shown in Fig. 2(a).

Thermal effects in MPA structure are studied at a small angle of incidence ($\mathrm{20^{0}}$) with laser pulses of 600ps pulse width and 1064 nm from a Q-switched pulsed laser (Wedge HF, Bright Solutions) with pulse repetition rates (PRR) between 3 to 10 kHz. Due to literally no transmission of the structure for 1064 nm, the experimental measurements are performed in the reflection configuration by focusing the incident laser beam with 10 cm focal length plano-convex lens and the reflected beam is collected onto a photo-diode. The experimental observations of the reflected pulse fluence as a function of the incident pulse fluence (monitored by a reference photo-diode) is plotted in Fig. 2(b). A nonlinear saturation in the reflected fluence is observed for various PRRs (3kHz to 10kHz). The nonlinear saturation of the reflected fluence depends on the PRR, with lesser saturation noted for smaller PRR at the same pulse energy. The nonlinear reflectance, normalized to the linear reflectance at small pulse energies, is also plotted with respect to the incident average power for each PRR in Fig. 2(c). This shows that the reflectance begins reducing with increase in average incident power for all the PRR. The rate of reduction is larger for large PRR. Thus, a nonlinear increase in the absorbance of the MPA that appears to be principally determined by the average laser power incident is noted. 

To understand the above nonlinear absorbance, we first need to analyse the linear absorbance mechanisms in the system. Numerical simulations of the MPA structure were performed for the fabricated structural parameters using the commercial finite element package, $\mathrm{ COMSOL^{\circledR}}$ Multiphysics~\cite{Comsol}. In the simulations, the temperature dependent dielectric permittivity values for Al films were considered from Hutnner et.al.~\cite{Huttner-Al} The refractive index of ZnS~\cite{Density-ZnS} is considered as $ n=2.75$ and non-dispersive within the bandwidth considered. A normally incident plane wave ($\theta=0^{0}$, Fig. 1(a)) on the MPA was considered. Perfect electric and magnetic conductor (PEC and PMC) boundary conditions are applied along the X and Y directions respectively for the incident TE and TM polarizations. The simulated reflectance spectrum obtained from the S-parameters of the finite element simulations, shows an absorbance band at 1090 nm along with another resonance band centred at 960 nm. This explains the absorption maximum $(\thicksim98\%)$ noted in the samples at 1090 nm. The two theoretical separate resonances are not visible in the measurements, while an asymmetry can still be discerned indicating the presence of more than one resonance within the broad absorption band. The experimental observations indicate a large level of inhomogeneous broadening, presumably due to the fabricational inaccuracies and limitations.

To understand the nature of the resonances, we calculate the electromagnetic fields excited in the MPA. The normalized electric field excited at the top Al metallic disk (Fig. 4(a)) indicates the excitation of the third order cavity-like mode supported by an antenna with an optical length of $ m\lambda/2=n_{Al}.(2r) $ for $(m=3)$.~\cite{GD-JOP} The magnetic resonance at 1090 nm and its normalized magnetic field confinement (Fig. 4(b)) in the middle dielectric ZnS layer results from the anti-parallel currents excited in the top and bottom metallic layers (white arrows in Fig. 4(b)). The simultaneous excitation of the electric and the magnetic resonances at 1090 nm results in the absorbance band shown in Fig.2(a). Moreover, a propagating surface plasmon polariton mode supported by the periodic disk array and the ground plane at 960 nm is suggested by the field distribution simulations shown in Fig.3(a). The broad band in the experimental reflection spectrum (Fig.2(a)) arises from the overlap of these resonances at 1090 nm and 960 nm due to inhomogeneous broadening. The first order dipole resonance of the MPA would be present at 4.5 $\mu$m (not shown here). The higher order resonance corresponding to $5^{th}$ order can be noticed at 675 nm in the simulated spectrum. 

\begin{figure}
\includegraphics[scale=0.17]{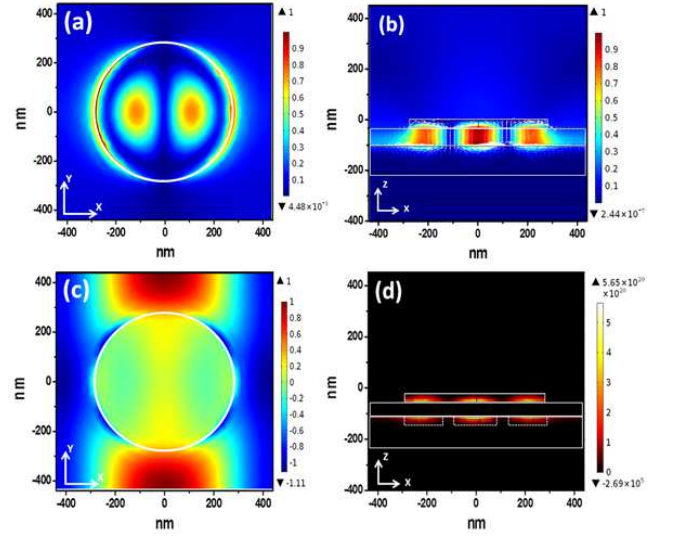}
\caption{ (a)Electric field distribution in the top Al disk and (b)magnetic field distribution in the spacer ZnS layer for the $\mathrm{3^{rd}}$ order magnetic resonance at 1090 nm. The displacement current density in the top and bottom Al layers are shown by the white arrows. (c) Electric field distribution at the top MPA surface for propagating plasmon mode at 960 nm. (d)Simulated heat dissipation in the top Al disk and ground Al films for the $\mathrm{3^{rd}}$ order resonance at 1090 nm.}
\end{figure}

The large absorption of radiation results in resistive/ohmic heating and elevated temperatures of the MPA. The intense localized resistive heating in Al components of the structure is shown in Fig.4(d) by simulating the total heat dissipation in the PA. Fig. 4(d) shows that there are localized hot regions symmetrically located on both the disk and regions close to the disk on the ground plane corresponding to the $3^{rd}$ order mode. The elevated temperatures can modify the optical constants of the heated metallic components, thereby change the spectral response of the MPA. A rigorous estimate of the transient elevated temperatures for given laser intensities and PRRs is required to obtain the material parameters at the incident laser powers.

We note that the peak temperature $\mathrm{(T_{high})}$ in the structure will depend on the pulse energy focused into the MPA while the lowest temperature $\mathrm{(T_{low})}$, that the structure will relax to, will depend on the PRR. Each pulse will encounter the material parameters at $\mathrm{T_{low}}$ as the laser pulse width is much smaller than the thermal relaxation rates and the inter-pulse period. For an incident pulse energy Q, we have $Q=C\mathrm{(T_{high}-T_{low})}$, where C is the heat capacity of the MTM in the focused area given by $C=m_{Al}C_{Al}+m_{ZnS}C_{ZnS}$, where m and C stand for the mass and specific heat of the material components. $\mathrm{T_{low}}$, is given by $ \mathrm{T_{low}= T_{high} exp (-\tau_p/\tau_r)}$, where $\mathrm{\tau_r=(\rho_{Al}C_{Al})R^2/k_{Al}}$ is the relaxation time~\cite{Boyd} and $\mathrm{\tau_p}$ is the inter-pulse period. Here $\mathrm{\rho}$ is the density, C is the specific heat, k is the thermal conductivity and R is the radius of the beam ($\mathrm{40 \mu m}$) at the MTM. We take $\rho=\mathrm{2.50 g/cm^3,3.70 g/cm^3}, C=\mathrm{0.9 J/gm-K, 0.47 J/gm-K}$ and $k=\mathrm{2.05 W/cm-K, 0.27 W/cm-K}$ for aluminium~\cite{Huttner-Al} and ZnS~\cite{Density-ZnS} respectively. The contribution of ZnS layer to heat flow is neglected due to the much smaller thermal conductivity. Thus we obtain for
\begin{equation}
T_{low}=\dfrac{Q}{C\mathrm{[\exp{(\tau_p/\tau_r)}-1]}}+T_0,
\end{equation}
where $\mathrm{T_0}$ is the room temperature (300 K). The $\mathrm{T_{low}}$ for different PRRs and same average laser powers are plotted in Fig. 3(b). Note that the higher PRR result in a large increase of temperature ($\mathrm{T_{low}}$) for 10kHz. The optical constants n and k~\cite{Huttner-Al} of Al for different temperatures are tabulated in Table I. 
\begin{table}
\caption{\label{tab:table1} n and k of Al at $\mathrm{\lambda=1064 nm}$ and different temperatures.~\cite{Huttner-Al}}
\begin{ruledtabular}
\begin{tabular}{lcr}
Temp(K)&$n(\pm0.005)$&$k(\pm0.005)$\\
\hline
300 & 1.589 & 10.334\\
360 & 1.726 & 10.533\\
420 & 1.837 & 10.807\\
450 & 1.893 & 10.923\\
480 & 1.960 & 11.055\\
\end{tabular}
\end{ruledtabular}
\end{table}
A maximum temperature change of 177K from room temperature can be obtained at 10kHz PRR for a pulse energy of $\mathrm{0.77 \mu J}$. The respective changes in the real and imaginary parts of the refractive index of Al were included in the finite element simulations to include the spectral responses at the different $\mathrm{T_{low}}$. The thermally induced increase in optical constants (n and k) values results a blue shift of the absorption maximum (Fig.3(c)) from 1090 nm at 300 K to 1064 nm at 477 K. As the resonance maximum approaches the excitation wavelength, a saturation in the reflectance is noticeable from the Fig. 3(c), with an increased absorption of the incident radiation at 1064 nm due to the shift in the resonance wavelength with increasing temperature. In case of lower PRR, the increase in the temperatures are much lower than for 10kHz. The blue shift of the resonance with respect to the increased average laser power is clearly indicative of a resonance mediated nonlinear absorption in the MPAs. This saturable reflectance or reverse saturable absorption observed for different PRRs and similar pulse energies originate from the photo-thermal changes induced in the MPA. The nonlinear absorption does not depend on the peak laser intensities of the short pulses, but rather on the average laser power. The spectral shifts induced by these photo-thermal effects can result in either reverse saturable absorption or saturable absorption depending on whether the laser is blue or red-shifted with respect to the MTM resonance.

In summary, we have demonstrated thermally induced resonance mediated nonlinear optical absorbance in a MPA. The intense and localized absorption of laser radiation result in resistive heating of the MPA. The thermally modified dielectric permittivity of Al is responsible for a dynamical tuning of the MPA resonance. Moreover, our fabrication technique using laser-interference lithography shows great potential for the manufacture of sub-micron sized MTM unit cells for operation at visible and NIR frequencies. Our pulsed laser measurements of nonlinear absorption in MPA also demonstrates the ruggedness of the structures, which are able to withstand peak pulse intensities exceeding $\mathrm{25MW/cm^{2}}$ without any damage. 

This work was supported by DRDO, India under grant no. $\mathrm{DECS/15/15124/D(R\&D)/CARS-1}$. SG thanks the IIT Kanpur for a fellowship. RK thanks the Council for Scientific and Industrial Research, India, for a fellowship. Authors acknowledge Prof. Goutam Deo, IIT Kanpur for the NIR spectrometric measurements.

\end{document}